\documentclass[12pt]{article} 
\usepackage[a4paper]{typearea}
\typearea{0}

\usepackage{helvet}
\usepackage{graphicx}
\usepackage{amsmath}
\usepackage{times}
\usepackage {graphics,pstricks}
\usepackage[T1]{fontenc}
\usepackage[latin1]{inputenc}
\usepackage {pst-all}
\usepackage {multido}
\usepackage{color}

\usepackage{natbib}

\begin{document}

\begin{center}
{\Large \bf How can the green sulfur bacteria use quantum computing for light harvesting?}

\hspace{1cm}

{\large \em D Drakova$^1$ and G Doyen$^2$}\\
\hspace{.5cm}

{$^1$ State University of Sofia, Faculty of Chemistry, Sofia, Bulgaria} \\
{$^2$ Ludwig-Maximilians-University, Munich, Germany}
\end{center}

\begin{abstract}
Long lasting coherence in photosynthetic pigment-protein complexes
has been observed even at physiological temperatures \cite{engeletal1,engeletal,Pani2011}.
Experiments have demonstrated quantum coherent behaviour
in the long-time operation of the D-Wave quantum computer as well \cite{johnson,dickson}.
Quantum coherence is the common feature between the two phenomena.
However, the 'decoherence time' of a single
flux qubit, the component of the D-Wave quantum computer,
is reported to be on the order of nanoseconds, which
is comparable to the time for
a single operation and much shorter than the time required to carry out
a computation on the order of seconds. An explanations for the factor of
$10^8$ discrepancy between the single flux qubit coherence time and the long-time quantum
behaviour of an array of thousand flux qubits was
suggested within a theory where the flux qubits are
coupled to an environment of particles
called gravonons of high density of states
\cite{arxiv,dice2014}. The coherent evolution is in high dimensional
spacetime and can be understood as a solution of Schr\"odinger's
time-dependent equation. \\

\noindent Explanations for the quantum beats observed in
2D Fourier transform electronic spectroscopy of the Fenna-Matthews-Olson (FMO) protein complex
in the green sulfur bacteria
are presently sought in constructing transport theories based on quantum master equations
where 'good' molecular vibrations ('coloured noise') in the chlorophyll
and the surrounding protein scaffold
knock the exciton oscillations back into coherence \cite{mohseni1,mohseni}. These 'good'
vibrations are claimed to have developed in three billion years of natural selection. These
theories, however, face the discomforting experimental observation that
"attempts to scramble vibrational modes or to shift resonances with
isotopic substitution miserably failed to affect the beating signals" \cite{gsengel}. As a possible way
out of this dilemma we adopted the formalism of the quantum computation to the quantum beats
in the FMO protein complex.\\

\end{abstract}

\section {Introduction}
\noindent When in 2007 the New York Times published an article suggesting that
plants were quantum computers, quantum information scientists exploded
into laughter \cite{khalili}.
Their concern was, of course, decoherence which is believed to be one
of the fastest and most efficient processes in nature. The laughter
is over and quantum biology is firmly established, but the mechanism
which sets decoherence apart is still under debate. \\

\noindent
\noindent The experiments by Engel et al.  show
that long lasting coherence in photosynthetic pigment-protein complexes
is observed at physiological temperatures \cite{engeletal1,engeletal,Pani2011}.
Quantum coherence over hundreds of femtoseconds in the initial steps of
photosynthesis, the tranfer of light quanta towards the reaction centers,
in green and purple bacteria \cite{lee,engel-vibronic} at room temperature \cite{harel,fidler1},
algae at ambient temperature of 294 K \cite{algae}
and even in living cells of purple bacterium \cite{dahlberg,Engel-livecell}
warrant the high efficiency of the
light-harvesting process. It has even been suggested
that the high efficiency of photosynthesis may be associated
with coherent energy and charge transfer on a long timescale
\cite{rebentrost,mohseni2,ishizaki}. \\

\noindent Citing Hildner et al.:
''...quantum coherence is the central aspect connecting
fields as diverse as quantum computing and light-harvesting'' \cite{hildner}. \\

\noindent In 2011 the first commercially available quantum computer
was presented by the company D-Wave
consisting of an array of more than thousand flux qubits (June, 2015),
which solves optimization problems
with many free parameters \cite{johnson,dickson}.
It works as a quantum computer of the quantum annealing type
for minutes and even for half an hour. However, a single flux qubit
in a Ramsey interference experiment shows coherent oscillations between the two supercurrents
for time of the order of 20 ns, which is
8 orders of magnitude shorter than the coherence time of the D-Wave computer \cite{chiorescu}.
The amplitude of the oscillations
decays to $1/{\rm e}$ after 20-25 ns. It is often said that
the flux qubit loses coherence within 20 ns, or it dephases, or it is decohered, or it is driven by noise
or experimental imperfections. The experimental observation is that
the probability for the oscillatory state change decreases exponentially,
achieving some non-zero final value,
which does not change with time any more.\\

\noindent This was a noteworthy result as a flux qubit is a macroscopic device
with dimensions of the order of $10^4$ {\rm \AA}. But not less
noteworthy was the experimental observation of a coherent quantum
behaviour of the Fenna-Matthews-Olson pigment-protein complex
not only at low temperature but even at 277 K and the coherent behaviour
of the light harvesting complex in the purple bacterium living cell in a dense, wet
noisy environment \cite{dahlberg,Engel-livecell}.
The 'time of coherent development' of the FMO-complex is picoseconds, i.e.
4 orders of magnitude shorter than that of a single flux qubit.
With the help of 2D Fourier transfrom electronic spectroscopy \cite{hybl}-\cite{abramavicius}
the amplitude of a photon
echo signal as a function of the waiting time between the first two laser pulses,
which excite the FMO electronically,
and the third laser pulse, which stimulates the photon emission from the exciton, shows beats.
The beats, the oscillations in the time development of the photon echo signal,
are due to the coherent superposition of two lowest energy
excitons shared by all chlorophyll molecules and indicate coherent time development of the FMO-complex
during the waiting time.
In contrast to the experimental observation, an exponential decay
of the photon signal with time would be expected,
if decoherence due to interaction with the noisy environment (phonons)
of the FMO-complex were effective, as it is generally
believed that electronic coherence decays on 10-100 fs timescales
\cite{nagy,prezhdo}. \\

\noindent Why is the beating between exciton states observed for
picoseconds instead of a fast exponentially decaying signal without oscillations,
as decoherence theory suggests?
Why is no decay to zero amplitude of the photon signal observed,
if decoherence due to interaction with the noisy environment (phonons)
of the FMO-complex were effective?
The expected question is: why does decoherence not take over
and reduce the photon signal of the FMO complex, which is in an environment of
many degrees of freedom at high temperature?
Decoherence theory aims to calculate the time development
of the reduced density matrix of a subsystem and thereby tries
to demonstrate that the reduced density matrix very rapidly
approaches a quasi-diagonal form. The diagonal matrix elements
are then interpreted as probabilities of finding the subsystem in the respective state.
Calculations for realistic models are scarce. Notable are those
by Joos and Zeh \cite{JoosZeh},
Zurek \cite{zurek} and, applying to the present investigation, by Lloyd et al. \cite{shabani}. \\

\noindent In contrast to expectations based on the decoherence concept,
both the molecular system FMO and the macroscopic flux qubit
show long time coherence. The major question is:
Why does decoherence in both cases play no role?
The damping of the Ramsey fringes in the experiment with the single flux qubit
cannot be due to decoherence, because otherwise the D-Wave quantum computer,
with an array of 1000 flux qubits,
would not be able to function as a quantum computer for minutes.  \\

\noindent Semi-classical diffusive models based on stochastic dynamics
have been extensively applied to study energy and charge transport in
biological systems \cite{foerster,marcus,hopfield}.
Within a quantum master equation approach the
time development of the reduced density matrix is evaluated \cite{shabani},
the effect of the
phonon environment is simulated by an additive Redfield term, which is dropped
for the study of the time development.
With a very special selection of the phonon environment (coloured noise)
the beats of the photon amplitude are reproduced, but the finite final value of the
photon amplitude is not explained. Furthermore the 2D Fourier transform
experiment shows that
no special phonons are involved in the damping of the signal,
changing the FMO phonon spectrum does not influence the results \cite{gsengel}.\\

\noindent Theoretical attempts based on the electron incoherent hopping process \cite{foerster,dexter}
between pigment molecules in the photosynthetic unit have been
suggested \cite{shulten}. \\

\noindent A discussion topic is the nature of the superimposed coherent states
in the protein-pigment complexes.
Vibrational, electronic or vibronic states may be involved,
with indications from experiment favouring their vibronic character
(\cite{engel-vibronic}, \cite{chenu}-\cite{womick}). \\

\noindent We focus on the experimental observation in the 2D Fourier transform electronic spectroscopy
which demonstrates coherence
by the Fenna-Matthews-Olson protein-chlorophyll
complex of green sulfur bacteria over times longer than 1000 fs and
even at high temperature (277 K). The amplitude of the measured photon echo signal
oscillates with time, which is an indication of coherent behaviour,
suggested as the basis for the extreme
efficiency of the photon transfer towards the reaction center and hence of the
photosynthetic processes in the depth of the Black sea.
The expectation that decoherence due to the chaotic molecular environment,
considered to be very fast, on the timescale of $10^{-20}-10^{-17}$ s,
in the bacterium cell at physiological temperatures will destroy coherence,
is not confirmed experimentally. The issues we will address in the present article refer to:
\begin{itemize}
\item Why is decoherence not effective and not fast enough to destroy coherence?
\item Despite that the photon amplitude beats are exponentially damped with time,
the final value is finite, larger than zero even at high temperature. What is the reason for this?
\item Missing isotope effect, random exchanges of $^1$H atoms or of $^{12}C$ with their isotopes
do not change the picture. Random hydrogen exchange with deuterium and exchange of
all $^{12}C$ with $^{13}C$ or even the whole hydrocarbon side chain of the pigment by a
different one do not affect the coherent behaviour \cite{engel-isotopes}.
This procedure certainly changes the phonon spectrum, but the results
on the coherent quantum beats are not influenced.
Coherence is experimentally shown in
bacteria \cite{engeletal1,engeletal,lee,fidler1,dahlberg,Engel-livecell}, algae \cite{algae}
and green plants \cite{schlau-cohen}.
The structure of their protein backbones
differ and hence the protein scaffolds cannot serve as the ubiquitous protection of
electronic coherence, as it is suggested in refs. \cite{engeletal1,engeletal,lee,algae}.
\item
A discusion issue is the nature of the coherent states and the role of the environment. Our
results are in favour of vibronic states, many-particle states involving
electron exciton components and local vibrations of the protein backbone.
\end{itemize}

\noindent
The dying out of the oscillations between the two supercurrents in the
single flux qubit was explained in a coherent picture solving Shr\"odinger's
time dependent equation in high dimensional spacetime \cite{dice2014}.
The decisive features of the theory are summarized in
the next section together with the description of
a simple model of the physical picture of the processes in FMO 2D Fourier transform electronic spectroscopy.
The results are presented in section \ref{results}, followed by a summary and conclusions. \\

\section{The model and the physical picture} \label{modelphyspicture}
\noindent The FMO protein complex consists of seven
chlorophyll molecules in a protein cavity.
There is an eighth bacteriochlorophyll molecule outside the cavity
whose interaction with the rest seven molecules can be neglected.
The first two laser pulses in the 2D spectroscopic experiment, delayed by only 20 fs from each other,
excite two excitons and lead to their coherent superposition, which results
in the beating signal.
\begin{figure}{} \begin{center} \begin{minipage}{15cm}
\scalebox{0.4}{\includegraphics*{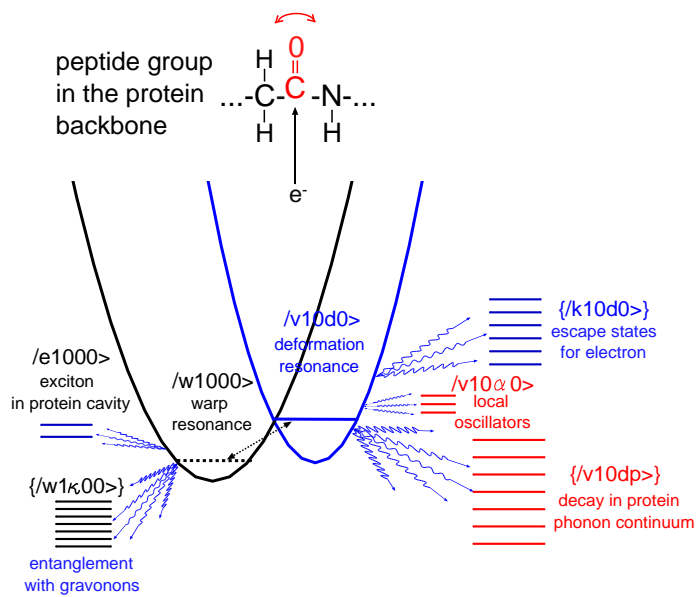}}
\end{minipage}
\end{center}
\caption{Model for studying the quantum coherent behaviour of a complex of bacteriochlorophylls (FMO complex)
in a protein scaffold in the green sulfur bacteria. The protein scaffold provides
the carbon atom in the carbonyl $CO$ fragment to capture the nuclephile, the exciton electron,
in a transient negative ion resonance which relaxes and couples with phonon modes in the protein
and excites local rocking and wagging motion of the carbonyl $CO$ fragment.
The entanglement with gravonons is effective within the NIR, but as it develops
slowly compared to the coupling with the phonons,
its effect during the tiny burst, induced by the double laser pulse, is neglected in the calculations.
In this schematic drawing the gravonon spectrum is plotted on a scale
magnified by a factor $10^8$ and differs from the scale of
the other spectra.
\label{twooscillators2}} \end{figure}
The time between the first two laser pulses,
i.e. the coherence time $\tau$, in different 2D spectroscopic experiments
is of the order of 20 to 500 fs
and the duration of each pulse is 15-40 fs.
These times are very short compared
to the time needed to establish the entanglement with the gravonons
(representing local version of the gravitons),
the environmental excitations
we take into account in addition to the excitations
considered in other theories like phonons and electron-hole pairs.
Therefore for the
purposes of our theory the first two laser pulses are regarded as one double pulse.
The third laser pulse is applied after a waiting time varying from 0 fs to 1000-1500 fs
and serves to stimulate the emission of the photon, which is recorded
as the photon echo after the rephasing time $t$. \\

\noindent The physical picture we describe is the following.
The exciton electron wave packet is diffuse, penetrating in the protein scaffold,
where as a nuclephile in a nucleophilic-like reaction the exciton electron attacks
the positively polarized carbon atom of the carbonyl $CO$ fragment.
The electron can be transiently accomodated in the $3s$-affinity like orbital on the carbonyl carbon atom
creating a transient negative ion resonance (NIR).
Thus the electron penetrates in the carbon core region where it interacts gravitationally with the atom core.
At interatomic distances the strength of the gravitational interaction is enhanced
by many orders of magnitude, if hidden extra dimensions exist.
In the spirit of Einstein's general relativity
the gravitational interaction between masses deforms spacetime and we introduce
a basis state called warp resonance for the electron (fig. \ref{twooscillators2}),
where it entangles with the gravitons
(the messenger particles of the gravitational field).
Locally a soft gravonon structure is generated, which is constructed as a
correlated motion of 3-4 atoms leading to the warping.
The soft gravonons in our theory are the solution of interacting quantum harmonic oscillators
centered on the 3-4 atoms close to the NIR. For more details on the theory
concerning the generation of the gravonons see refs. \cite{dice2014,foundpaper}.\\

\noindent The electromagnetic field acts as an external force on the carbon atom
and, together with the thermal fluctuations, shifts it slightly in a different position
creating a local deformation in the protein chain.
This is no wonder since
the local vibrations of the atoms in the carbonyl $CO$ fragment are different if the carbon atom has
captured transiently the electron or not. The local deformation of the
protein chain occurs around the NIR, it is associated with slight increase in energy, and is
described by introducing a basis state called deformation resonance for the core motion of
the carbonyl fragment. Furthermore excitation of
vibrational modes like the rocking and wagging motion of the
carbonyl $CO$ fragment is possible which we describe by coupling the deformation resonance to
two local harmonic oscillators, i.e. to local vibrational modes (fig. \ref{twooscillators2}). \\

\noindent In the deformation resonance coupling to the extended phonons of the protein backbone
is also effective, leading to phonon excitation and energy dissipation from the local region
(fig. \ref{twooscillators2}).
The exciton electron has the option to decay from its localized state in the NIR in delocalized
electron states, called 'escape states' because total energy has to be conserved,
if phonons are excited. 
Electron states delocalized over the chlorophyll assembly with
the $3s$-affinity like orbital of the carbonyl carbon atom as component play this role. 
It is within the deformation resonance that coupling
with the phonons of the protein backbone is taken into account and energy dissipation
from the local region in the protein backbone via phonon excitation may occur.
However, due to the interaction between the
deformation resonance and the local rocking and wagging modes of the carbonyl $CO$ fragment,
reflection to and fro between the local states occurs. Therefore, despite the dissipation
into the protein backbone phonons, there remains a local non-zero electron component
in the total wave packet, which explains the finite non-zero value
of the photon signal in the 2D spectroscopic experiment when the deexcitation of the exciton electron is
stimulated by the third laser pulse.\\

\noindent
The gravitational interaction of the exciton electron with the carbon atom core
has to be in high spacetime dimensions ($11D$) as string theory
assumes. Only then gravity is strong enough at small distances and decays fast with distance
($r^{-8}$ power law and nearly 33 orders of magnitude higher value of the gravitational constant in $11D$).
The additional 7 spacial dimensions are compactified and are hidden,
so that there is no discrepancy with Newton's gravitational law at distances where it was proved valid.
Within the warped space the time development of the
entanglement of the electron with gravonons
leads to the beables.\\

\noindent
Beables are mathematically precisely defined as a set of configurations
in the expansion of the wave function containing a localized matter field
and excited local gravonons.
Loosely speaking beables
are matter fields (e.g. atoms or electrons)  in $3D$ space, entangled
with gravonons, which interact with
other atoms gravitationally and generate the gravonon structure in high
dimensional spacetime.
Therefore localized atoms, molecules or electrons in $3D$ space
entangled with gravonons in high dimensions in the form of beables
exist as long as the beables and the entanglement with the gravonons
exist.\\ 

\noindent
The word "beable" has been coined by John Bell \cite{bell}
as a terminology against the word observable: "The concept of
'observable' lends itself to very precise mathematics when identified
with 'self-adjoined operator'. But physically, it is a rather wooly
concept. It is not easy to identify precisely which physical processes
are to be given the status of 'observations' and which are to be
relegated to the limbo between one observation and another.'' \\

\noindent
We use the expression beable in the sense of John Bell's {\em local}
beable. According to our proposal
signals in experiments in $3D$ space can be received only via beables
and hence measurements are tied to beables \cite{foundpaper}.
Expressed with John Bell's words: "One of the apparent non-localities
of quantum mechanics is the instantaneous, over all space, 'collapse
of the wave function' on 'measurement'. But this does not bother us if
we do not grant beable status to the wave function."\\

\noindent
Already in 1927 it was revealed that an interference pattern shows up
on a photographic plate only when the number of photons falling on the
plate is very large \cite{dempter-batho}.
The history of photon detection is nicely reviewed in ref. \cite{dempkhalili}.
Here the results of Dempter and Batho are summarized in such a way that
when, during
the so called 'collapse of the wave function', the photon is destroyed,
there appears somewhere on the photographic plate an atom of elemental silver
which will act as an embryo from which, by photographic development, a small
seed of silver will grow. The silver embryo is much smaller than the electromagnetic
wavelength and constitutes the beable in our picture.\\

\noindent
Applied to the spectroscopic method used in the bacteria experiment inestigated here
this means that the photon of the laser pulse occupies during the interference
process at least the whole volume of the
cavity embraced by the protein scaffold, but if the photon is measured and hence destroyed,
there appears a beable (embryo) formed in the detection devices of the experimenters.\\

\noindent
The described
mechanism of the so called collapse of the wave function has by now
also been established for matter fields \cite{arndt1}-\cite{arndt3}. For an electron in
an excitonic state of the chlorophyll it means that the wave function
of this electron occupies during the interference process the whole volume
of the chlorophyll, but when it hits the protein scaffold,
which acts as a kind of screen, the electron becomes localized in a tiny volume
of the size of the silver embryo, this being the carbon affinity level of the
carbonyl group in the model developed here. \\

\noindent 
Before the laser pulse is sent in, the electron will be almost with certainty
in such a beable. The laser pulse initiates then an interference process
where the many particle wave function describing photon, electron,
phonon etc. extends over the whole volume embraced by the protein scaffold.
This many particle wave function is not a beable, but, in John Bell's
terminology, rather a 'limbo state'. As time passes on a matter field will
somewhere entangle with gravonons and form a beable which 
destroys the photon. If beforehand the photon 'localizes' in a beable
in the detection devices of the experimenters, it will do so proportional
to the light intensity in the system.

\subsection {Chooser mechanism, beables created with the chooser mechanism}
\noindent The question is why does the exciton electron become
localized on a single carbonyl carbon atom of a single carbonyl $CO$ fragment in the protein backbone?
Quantum particles, e.g. the exciton electron,
may become localized via the entanglement with gravonons in the form of beables.
However, as the gravitational interaction even in high dimensional
spacetime is weak, entanglement with gravonons can be effective on the energy shell alone.
In the present model the exciton electron is localized
because via the entanglement to gravonons a beable is generated
by the local and strongly distant dependent interaction with the
gravonons. In the case of on-shell coupling even a very weak coupling
with the gravonons is effective. Just a few or a single carbonyl carbon atom in the protein
chain provide the condition for on-shell coupling with the initial exciton electron wave packet.
This is the chooser mechanism for the
site of most favourable entanglement with gravonons.
The chooser mechanism is responsible for the appearance of stuck
particles on single sites on the detection screen in double slit experiments \cite{doubleslit}.\\

\noindent The exciton is a diffuse object. Its overlap with all carbonyl $CO$ fragments
in the protein backbone would mean that the electron might be thought
to reside on all carbonyl carbon atoms in $CO$ sites at the same time and
would couple to all degrees of freedom all over the protein chain.
Then decay in the environmental degrees of freedom
resembles decoherence and is commonly predicted to be very fast.
This is, however, not the case in our theory
because the chooser mechanism selects a single carbonyl carbon atom on a single $CO$ site where
the condition for degeneracy coupling with the gravonons is fulfilled,
i.e. where the chooser mechanism leading to the transient localization of the electron works.
The electron can go to only a single carbonyl carbon atom in a single $CO$ site,
chosen by the degeneracy coupling criterion.\\

\begin{figure}{} \begin{center} \begin{minipage}{15cm}
\scalebox{0.4}{\includegraphics*{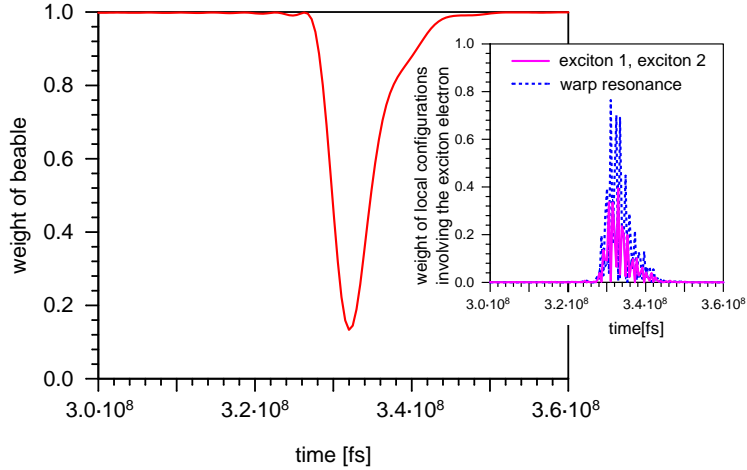}}
\end{minipage}
\end{center}
\caption{Time development of the entanglement of a quantum particle, e.g.
the exciton electron, with the gravonons,
generating the beable. Within the beable the electron is localized in
the warp resonance and can interact with the photon field, i.e. it can be measured.
As soon as the lifetime of the gravonons in the
high hidden dimensions expires, the electron bursts out of the entanglement with the
gravonons and is released in $3D$ space. This is the tiny burst in the figure.
It exists for a short time before the entanglement with the gravonons is
reestablished, which takes $5\cdot 10^{-9}-10^{-8}$ s. The inset shows the
bursting of the exciton electron out of the beable in three dimensional space
(in the warp resonance and the excitons in the cavity) during the tiny burst.
\label{beable}} \end{figure}

\noindent Beables are destroyed when the entanglement with the gravonons
is changed or truncated. This occurs as a tiny burst in the time development of the beable
caused for instance by the laser pulses
(fig. \ref{beable}). In the tiny burst the quantum particles (atoms, molecules, electrons)
are released from the entanglement with the gravonons and they burst out of the beable into $3D$ space
in a 'limbo state'. All states where particles are not entangled to gravonons
we call limbo states. \\

\noindent
Within the beable the electron is localized in
the warp resonance and can be measured.
As the lifetime of the gravonons in the
high hidden dimensions expires, the entanglement
of the localized electron with the gravonons is truncated and the electron is released in
$3D$ space. The electron bursts out of the entanglement with the
gravonons. \\

\noindent To demonstrate the destruction and re-creation of the beable
the Hamiltonian, described in section \ref{totalhamilt}, is diagonalized
without the term $H_{el-grav}$. Then this term is transformed to the
new diagonalized basis and the time dependent Schr\"odinger
equation (TDSE) is solved as described in section \ref{totalhamilt}.
The tiny burst can be seen in fig. \ref{beable}.
It exists for a short time before the entanglement with the gravonons is
reestablished which takes $5\cdot 10^{-9}-10^{-8}$ s.
Within the tiny burst experimentally nothing can be recorded
about the electron. This situation does exist in different experiments:
in double slit diffraction experiments with electrons and molecular beams
we know nothing about the quantum particles during their time of flight between the
source and the detection screen; in laser-induced desorption we know
nothing about the adsorbate for the time of flight between the
state with the adsorbate as a beable on the substrate surface and in the mass spectrometer,
where a new beable is created.\\

\noindent The effect of the double laser pulse in experiment
is to induce a tiny burst in the time development
since the laser pulse creates an electromagnetic field,
which exerts a force on the localized particle within a beable.
Slight shifts of the atom positions mean changed local
gravonon structure and destroyed coupling to the gravonons
(very short ranged), hence the beable is destroyed.
For instance, the double laser pulse in the 2D spectroscopic experiment
interacting with the electric dipole due to the localized electron on the carbon atom,
induces a slight shift of the carbon atom
and rocking and wagging motion of the carbonyl $CO^-$ fragment in the protein. This would cause
such a drastical change in the gravonon structure, typical for the beable
configuration existing before the laser induced tiny burst, that the entanglement
with the gravonons is destroyed. This is how the
gravonon structure and the beable are destroyed by the laser pulses.
As the entanglement with the gravonons develops in a period of time of the order of $10^{-8}$ s,
in the laser-pulse-induced tiny burst the electron is released
from the entanglement with the gravonons, however, its
coupling with other excitations, living in 4 dimensional spacetime
like phonons and the vibrations of the local oscillators, gains weight.
Although experimentally an electron, a photon or a molecule
released from a beable cannot be directly detected during the tiny burst, we can
calculate the time development further with the time dependent Schr\"odinger equation.
This is where in our theory we start a new time development: the electron is localized
in a transient NIR on a carbonyl carbon atom in a single carbonyl
$CO$ fragment, the photon is in the protein cavity, no gravonons are involved.\\

\noindent During the tiny burst
the exciton electron is no more entangled with the gravonons, it is
released in a limbo state in $3D$ space,
therefore cannot be directly accessed experimentally.
But everything in the 2D spectroscopic experiment
occurs during the tiny burst induced by the double laser pulse.
This situation lasts untill the entanglement with gravonons can develop
again, giving rise to the beables. During the lifetime of the beable
the exciton electon, which initially is a delocalized electron wave packet, remains localized
transiently on the carbonyl carbon atom.
Within this physical picture all measurable phenomena
originate effectively from a very local region which then expands in $3D$ space.
The chooser mechanism for the generation of the beables
and the concept of the tiny burst due to the laser pulse
indicate that what is measured in experiment has to be interpreted in a local picture
concerning just a single
carbonyl $CO$ fragment in the protein backbone, which couples to no more than
two local harmonic oscillators and the phonon excitations in the protein backbone. If more
local harmonic oscillators were excited, which is possible only  in the absence of the
chooser mechanism, the result would be indeed fast
decoherence of the photon beating signal.
This is, however, not measured in experiment. \\

\section{ The Hamiltonian} \label{totalhamilt}
\noindent The quantum fields in the model are:
(i) electron;
(ii) photon;
(iii) gravonons;
(iv) core movement field of the carbonyl $CO$ fragment;
(v) phonons.
The total Hamiltonian, the gravonons taken into account as well, includes terms describing
these fields and the many-particle interactions with the electron field:
\begin{eqnarray}
H_{FMO}&=&H_{el}+H_{phot}+H_{grav}+H_{CO}+H_{phon} \nonumber \\
&+&H_{el-phot}+H_{el-grav}+H_{el-CO}+H_{el-phon}.
\end{eqnarray}
The Hamiltonian for the electron includes single electron terms
for exciton 1 within the chlorophyll assembly $\mid e \rangle$,
exciton 2 in the chlorophyll assembly $\mid e' \rangle$,
exciton 1 in the warp resonance $\mid w \rangle$,
electron in the deformation resonance $\mid v \rangle$,
electron in delocalized 'escape states' $\{ \mid k\rangle \}$,
the interaction of exciton 1 with the warp resonance and the interaction between the two excitons:
\begin{eqnarray} \label{electronhamilt}
H_{el}&=&E_ec_e^+c_e+E_{e'}c_{e'}^+c_{e'} + + E_{w}c_{w}^+c_{w}
+E_vc_v^+c_v+\sum_k E_kc_k^+c_k  \nonumber \\
&+& W_{e}(c_w^+c_{e}+c_{e}^+c_w) + W_{ee'}(c_e^+c_{e'}+c_{e'}^+c_e).
\end{eqnarray}
The Hamiltonian for the photon is:
\begin{equation}
H_{phot}=\omega_{phot}a^+a
\end{equation}
and for the non-perturbed gravonons:
\begin{equation}
H_{grav}=\sum_{\kappa}^{}\varepsilon_{\kappa}\zeta^+_{\kappa}\zeta_{\kappa}.
\end{equation}
The Hamiltonian for the core movement of the carbonyl $CO$ fragment in the deformation resonance
(initial state and excited states) includes also the two local oscillators, describing the
wagging and rocking motion of the carbonyl group, and their interaction:
\begin{eqnarray}
H_{CO}&=&\epsilon_od_o^+d_o+ E_dd^+d
+\sum_{i=1,2;n_i=0}^{\infty} n_i\omega_{i} d_{n_i}^+d_{n_i}  \nonumber \\
&+&W_d \sum_{i=1,2;n_i=0}^{\infty}(d^+d_{n_i}+d_{n_i}^+d).
\end{eqnarray}
The Hamiltonian for the phonons is:
\begin{equation}
H_{phon}=\sum_p\omega_{p}b^+_p b_p.
\end{equation}
The many particle electron - photon interaction, giving rise to an exciton polariton, is
described by the Hamiltonian:
\begin{equation}
H_{el-phot}=V_g(c_e^+c_{e'}+c_{e'}^+c_e)(a^++a)
\end{equation}
and the many particle electron - gravonon interaction is described by the Hamiltonian:
\begin{equation}
H_{el-grav}= Y_wc_w^+c_w \sum_{\kappa}(\zeta_{\kappa}^+ \zeta_o+\zeta_o^+\zeta_{\kappa}).
\end{equation}
The many particle interaction between the electron in the deformation resonance and
the core movement of the carbonyl group includes interaction
terms with the core movement states in the deformation resonance and in the local oscillators:
\begin{eqnarray} \label{h-el-co}
H_{el-CO}&=& V_{core-def}(c_w^+c_v +c_v^+c_w)(d_o^+d+d^+d_o) \nonumber \\
&+&V_{core-loc}c_v^+c_v\left( \sum_{i=1,2;n_i=0}^{\infty}d^+d_{n_i}+
\sum_{i=1,2;n_i=0}^{\infty}d_{n_i}^+d\right),
\end{eqnarray}
giving rise to the vibronic states.
The many particle electron - phonon interactions are contained in the term:
\begin{equation} \label{h-el-phonon}
H_{el-phon}= Y_p \sum_{k}(c_v^+c_k +c_k^+c_v)\sum_{p}(b_p^++b_p).
\end{equation}

\noindent The meaning of the symbols in eqs. (\ref{electronhamilt}-\ref{h-el-phonon}) is
as follows:\\
\noindent
$E_e, E_{e'}$ are single particle energies of exciton 1 and exciton 2 in the protein cavity and
$E_{w}, E_v, E_k$ denote the electron energy in the warp resonance $\mid w \rangle$
(localized at a single carbonyl carbon atom in one carbonyl $CO$ fragment in the protein backbone),
in the deformation resonance $\mid v \rangle$
and in the escape states $\{ \mid k \rangle \}$.
\noindent
$c_e^+,c_e,c_{e'}^+,c_{e'},c_w^+,c_w,c_v^+,c_v,c_k^+,c_k$ are the creation and destruction operators
for the electron in the respective single electron states. \\
\noindent
$W_e$ is the interaction strength between exciton 1 and the warp resonance;
$W_{ee'}$ is the interaction strength between exciton 1 and exciton 2.\\
\noindent
$\omega_{phot},a^+,a$ are used for the photon energy and photon creation and annihilation operators.  \\
\noindent
$\varepsilon_{\kappa},\zeta^+_{\kappa},\zeta_{\kappa}$ are the energy and
creation and annihilation operators
for the gravonons, which are treated in the quantum harmonic approximation.
$\omega_{grav}(\beta^+\beta + \frac{1}{2})\equiv \sum_\kappa \varepsilon_{\kappa}\zeta_{\kappa}^+\zeta_{\kappa}$
with $\varepsilon_{\kappa}=\omega_{grav}(n_{\kappa}+\frac{1}{2})$ is the parabola
representation of the gravonons.  \\
\noindent
$\epsilon_o, d^+_o,d_o$ are used for the energy, the
creation and annihilation operators for the initial core movement state
$\mid d_o \rangle$ of the carbonyl $CO$ fragment in the deformation resonance.
$E_d, d^+,d$ refer to the energy, creation and annihilation operators for the excited core movement state
$\mid d \rangle$ of the carbonyl $CO$ fragment in the deformation resonance.\\
\noindent
$n_i$ denotes the number of the energy level in the $i$-th local oscillator ($i=1,2$) with energy
$\omega_{i}$ and creation and annihiltion
operators $d^+_{n_i},d_{n_i}$  of vibrations in the $n$-th energy level of the $i$-th local oscillator.   \\
\noindent
$W_d$ is the coupling strength between the vibrations of the carbonyl group
$CO$ in the deformation resonance and the local oscillators.\\
\noindent
$\omega_{p}$ is the energy of the non-perturbed phonons in the protein backbone,
with $b^+_p,b_p$ the creation and annihilation operators for phonons in the protein backbone.\\
\noindent
$V_g$ is the electron-photon coupling strength in the cavity, creating the polariton.  \\
\noindent
$Y_{w}$ is the coupling strength between the electron in the warp resonance $\mid w \rangle$ and
the gravonons $\{ \mid \kappa \rangle \}$.\\
\noindent
With $V_{core-def}$ the interaction strength between the electron and the
core movement state of the carbonyl group $CO$
in the deformation resonance is denoted. \\
\noindent
$V_{core-loc}$ is the interaction strength between the electron in the deformation resonance
and the core movement states of the carbonyl group $CO$ in the local oscillators.\\
\noindent
$Y_{p}$ refers to the coupling strength between the electron and the phonons
$\{ \mid p \rangle \}$ in the protein backbone.\\

\noindent
The last term eq. (\ref{h-el-phonon}) shows that when a phonon is excited in the protein backbone
the electron has to decay from a local state in the deformation resonance
in a delocalized electron escape state. This warrants the energy conservation
required by Schr\"odinger's equation.  \\

\noindent The second term in eq. (\ref{h-el-co}) and the term in eq. (\ref{h-el-phonon})
are responsible for the quantum beats, their exponential damping
and the non-zero finite value of the photon signal
produced by the stimulated electron deexcitation.
The second term in $H_{el-CO}$ (eq. \ref{h-el-co}) describes
the interaction between
the electron in the deformation resonance of the carbonyl $CO$ fragment and
the local oscillators. The term eq. (\ref{h-el-phonon})
is the interaction of the electron with the delocalized phonons, excited in the protein backbone.
The second interaction term in eq. (\ref{h-el-co})
takes care of preserving the weight of local configurations
with the electron involved and is responsible for the finite final amplitude
of the photon signal, resulting from electron
deexcitation, stimulated by the 3rd laser pulse.
The interaction term eq. (\ref{h-el-phonon}) takes care
of the entanglement with the protein phonons. This is the dissipative term,
which leads to the attenuation of the weight of electron configurations in the
local region and causes the exponential damping of the amplitude of the
stimulated photon emission signal with time. \\

\noindent The method of solving Schr\"odinger's equation for
the time development of the total wave packet
has been described in ref. \cite{foundpaper}.
It uses a basis consiting of many particle configurations (the electron, the photon,
gravonons, core movement of the carbonyl $CO$ fragment, protein phonons)
in a configuration interaction (CI) approximation.
The local many particle basis includes
notations for all fields (with gravonons and phonons in their initial states). For instance:
$\mid e,1,0,0,0\rangle$, $\mid e',0,0,0,0\rangle$, $\mid w,1,0,0,0\rangle$, $\mid v,1,0,d,0\rangle$,
where in the first position the electron basis function is denoted,
in the second place the photon state,
in the third place: the gravonon state,
in the fourth place: core movement state of the carbonyl $CO$ fragment and
in the fifth place: the protein backbone phonon state.\\

\noindent The specific feature of the theoretical approach
applying to the green sulfur bacteria is the choice of the matrix elements.
The energy difference between exciton 1 and exciton 2 equals 22 meV
and is of the order of magnitude determined experimentally \cite{caram}
and used in other approaches \cite{ishizaki2}. For the interection matrix element between the two
excitons we use 20 meV, and 12 meV for the coupling between the exciton and
the warp resonance. \\

\noindent After diagonalizing to first order in the basis the time
dependent Schr\"odinger equation is solved:
\begin{equation}   \label{psioft}
\mid\Psi (t)\rangle = e^{-iH_{FMO}t}\mid \Psi(0)\rangle.
\end{equation}
As describen in section \ref{modelphyspicture} we solve the TDSE twice. In the first calculation
the initial state $\mid \Psi(0)\rangle$ is an extended exciton state
which develops into a localized beable. In the second calculation
the initial state $\mid \Psi(0)\rangle$ has the exciton electron in the beable state
($3s$-affinity orbital of the carbonyl carbon atom) as described in section \ref{modelphyspicture}. \\

\noindent The single particle basis states are:\\
\noindent (i) the electron in the excitons $\mid e \rangle$,
$\mid e' \rangle$ and in the warp resonance $\mid w \rangle$,
in the deformation resonance $\mid v \rangle$ and in the escape states $\{ \mid k \rangle\}$;\\
\noindent (ii) the photon: 1: with photon in the protein cavity, 0: no photon;\\
\noindent (iii) gravonons: 0: for initial state of the gravonon continuum, $\{ \kappa \}$ for excited gravonons;  \\
\noindent (iv) the carbonyl $CO$ group in its vibrational state in the deformation resonance:
(0 for initial vibrational state; $d$ in an excited vibrational state) and
the carbonyl $CO$ group in the wagging or rocking vibrational state,
described by the local oscillators
($\omega_i$ is the energy of the $i$-th local oscillator in its $n$-th state $\mid d_{n_i} \rangle$);\\
\noindent (v) the protein backbone phonon state: 0 for protein phonons in their initial state,
$\{ p \}$ for excited protein phonons. \\

\subsection {Vibrational energy in the local oscillators}
\noindent An estimate of the vibrational excitation energy of the rocking and wagging motions of the carbonyl
group in the protein backbone of the order of $\omega\approx 5$ meV
is based on the experimental data in ref. \cite{engeletal}.
With this estimate at thermal energy
of approximately $k_BT=11$ meV at $T=125$ K two local
vibrations can be excited each with $\omega_1=\omega_2 =\omega \approx 5$ meV,
so that $2\omega \approx 10$ meV corresponds approximately to the thermal energy at 125 K. 
The rocking and wagging motions of the carbonyl group are much softer
than its vibration with respect to the protein chain, hence 5 meV for the quantum of
these vibrations appears physically plausible. \\

\noindent The energy of a local vibration can be neither 4 meV nor 6 meV because these values
contradict the experimental curves for 77 K, 125 K and 150 K.
The estimate for the energy of the local mode
$\omega \approx 5$ meV, based on the 2D Fourier transform spectroscopy experiment, corresponds to
the thermal energy $k_BT$ at $T\approx 58$ K.
At $T=125$ K and $T=150$ K the two experimental curves coincide. If by
$T=125$ K two quanta are excited, then at $T=150$ K it is not possible to excite 3 quanta,
hence 2 excited quanta would explain the coinciding curves in experiment.
The two quanta $2\omega$ should correspond to $T=125$ K and
$k_BT\approx 11$ meV, hence $\omega \approx 5$ meV.
$\omega$ cannot be 4 meV ($T\approx 46$ K) because then at $T=150$ K three
vibrational quanta will be excited, whereas at $T=125$ K only two can be excited
and the two curves will not coincide.
$\omega$ cannot be 6 meV ($T\approx 70$ K) either because at $T=125$ K, $k_BT\approx 11$ meV,
just a single local vibration will be excited exactly as by $T=77$ K. The two
curves at $T=77$ K and $T=125$ K would coincide. This is, however, not observed experimentally. \\

\subsection {Protein phonon band}
\noindent The protein phonon band width is of the order of 0.03 eV.
This follows from the following argument. Assume a periodic constant of
a simple one dimensional protein of the order of $a\approx 10$ bohr, then the maximal
wave vector $k_{max}$
is of the order of $k_{max}=\frac{\pi}{a}\approx 0.3$ bohr$^{-1}$. The energy corresponding to
the maximal $k$-vector is $\omega_{p,max}=vk_{max}=0.027$ eV
(using for the sound velocity in dry air $v=330$ m/s$=6.6 \times 10^{12}$ bohr/s).
This estimate refers to the transversal phonon modes.  \\

\section {Results for the time development of the tiny burst} \label{results}
\noindent We assume that the exciton electron is in a beable before
the laser pulses occur, because the bacterium exists in reality and therefore
cannot be in a limbo state. Reality consists of beables only.\\

\noindent The results of the calculation leading to the beable have been
presented in fig. \ref{beable}. We now discuss what happens after the
beable has been destroyed by the laser pulse.\\

\begin{figure}{} \begin{center} \begin{minipage}{15cm}
\scalebox{0.4}{\includegraphics*{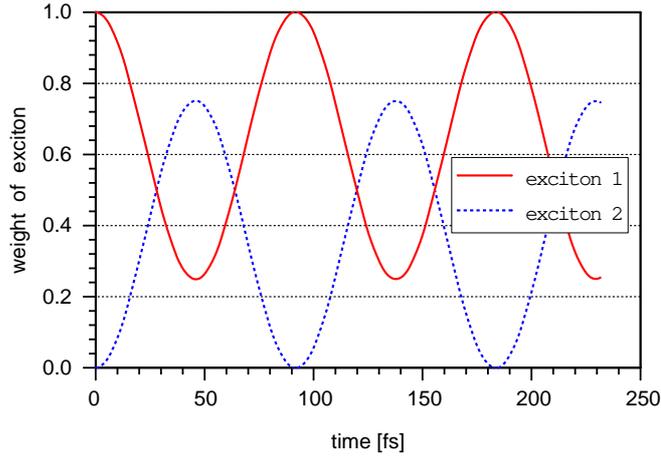}}
\end{minipage}
\end{center}
\caption{Rabi oscillations between the two exciton states
isolated from the environment.
\label{g1-e0}} \end{figure}

\noindent The first laser pulse in the 2D spectroscopic experiment excites
the FMO complex electronically producing the first exciton delocalized in the protein cavity.
The second laser pulse excites the second exciton and a coherent
superposition between the two \cite{fleming}.
If no interaction with the environment exists the coherent oscillations between
exciton 1 and exciton 2 will continue, the amplitude of each state changing
sinusoidally with time (fig. \ref{g1-e0}).\\

\begin{figure}{} \begin{center} \begin{minipage}{15cm}
\scalebox{0.35}{\includegraphics*{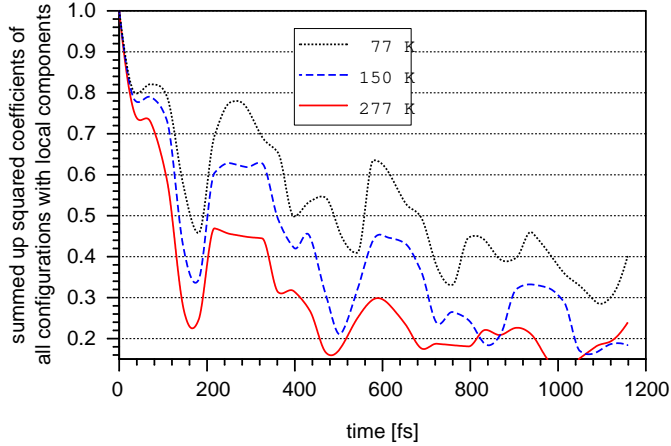}}
\end{minipage}
\end{center}
\caption{Time development of the amplitude of the photon signal during
the tiny burst, induced by the double laser pulse, for the model in fig. \ref{twooscillators2}
at three different temperatures of the environment. The photon signal scales
with the summed weight of the field configurations with
local components in the wave packet. These
are configurations with the exciton electron in the warp resonance
(i.e. in the beable), in the deformation
resonance and the local oscillators. The initial state is the beable
described in the text.
\label{temper-2osc}} \end{figure}

\noindent However, for the  FMO-protein complex we have the protein cavity
and the chlorophyll molecules within the cavity. Before the tiny burst the exciton
electron is transiently localized in a single warp resonance on a carbon atom
at a single carbonyl $CO$ fragment chosen
by degeneracy with the gravonons. The photon is in the cavity. The localized electron
exercises a force on the carbon atom, exciting two local oscillators with the wagging
and rocking movements of the carbonyl $CO$ fragment in the deformation resonance.
In addition to coupling to the local oscillators, coupling to the extended
protein phonons within the deformation resonance allows
relaxation and dissipation in delocalized phonons away from the local site.
In the 2D Fourier transform electronic spectroscopy experiments
these components are not observed, the measurement provides just the
amplitude of the photon echo, which scales with the weight of the local configurations involving
the exciton electron, as a function of time. But we can calculate
the time development of the wave packet during the period of the tiny burst and it
shows a strong similarity with the variations with time of the photon amplitude
measured experimentally (fig. \ref{temper-2osc}).\\

\noindent In experiment the third laser pulse after a waiting time $T$ stimulates
the photon emission. At time $T$ the photon is emitted with the amplitude,
with which the electron to be deexcited is represented in the local components of the wave packet
(warp resonance, deformation resonance, local oscillators) just before the third
laser pulse. Our theory ends at the moment just before the third laser pulse is shot.
The third laser pulse will cause stimulated photon emission
with the amplitude, with which the exciton electron is represented in all configurions involving the local
states: the warp resonance, the deformation resonance and the local oscillators.
The sum of the squared amplitudes of these configurations is plotted
as a function of time in fig. \ref{temper-2osc}.
The amplitude of the emitted photon signal scales with the weight of
all local configurations involving the exciton electron.
In experiment choosing the photon in the detector via a new beable happens
with an amplitude equal to the sum of the squared coefficients of those configurations, which have
the exciton electron, whose deexcitation is stimulated by the third laser pulse.

\subsection{Temperature effects}
\noindent For the solution of the time dependent
Schr\"odinger equation at different temperatures
we explicitely need the dependence of the accessible vibrational
configurations on temperature, i.e.
counting of the vibrational configurations in the deformation resonance,
and the temperature dependence of the accessible on-shell phonon configurations.
The assumption is that the deformation resonance is
as highly excited as the external temperature allows. This energy can then be
redistributed betwen the local channels and the delocalized phonons.\\

\noindent We assume that within the deformation resonance
coupling to two local quantum harmonic oscillators is effective,
associated with the rocking and wagging motions of the carbonyl $CO$ fragment.
Einstein's model is assumed (non-interacting oscillators
with the same frequency).
One local oscillator is not enough to reproduce the temperature dependence of the signal,
it is not able to compete with the dissipative effect of the protein phonons.
If it were just one local harmonic oscillator available for relaxation
(one relaxed vibrational mode) we would get fast decay of the photon amplitude and no temperature
dependence since one oscillator provides one local vibrational
configuration for all experimental temperatures.
If the number of coupled local oscillators and
local vibrational configurations were
large, i.e. the exciton electron resides transiently
on many carbonyl carbon atoms in many carbonyl $CO$ fragments,
we would get no temperature dependence of the decay of the photon
amplitude either. It would immediately drop to zero as decoherence theory suggests.
No chooser effect and fast decay in the protein phonons all over the protein chain would
be the result. This is
a situation when the local structure is lost.
Hence, the result would be fast exponential decay of the photon signal as
anticipated by decoherence theory.
However, in our model, starting from a delocalized exciton electron,
with the chooser mechanism due to on-shell entanglement with gravonons,
a localized electron in the form of a beable is generated.
The local nature of the interactions in this many particle system prevents
the excitation of many harmonic oscillators over the protein chain
and the total decay of the photon signal. \\

\noindent With two local oscillators we get the exponential damping
of the photon amplitude with time,
the beating of the photon signal and the finite non-zero value of the photon
signal at long time as shown in fig. \ref{temper-2osc}. The temperature dependence is also
in satisfactory agreement with the experimental observation.\\

\begin{figure}{} \begin{center} \begin{minipage}{15cm}
\scalebox{0.37}{\includegraphics*{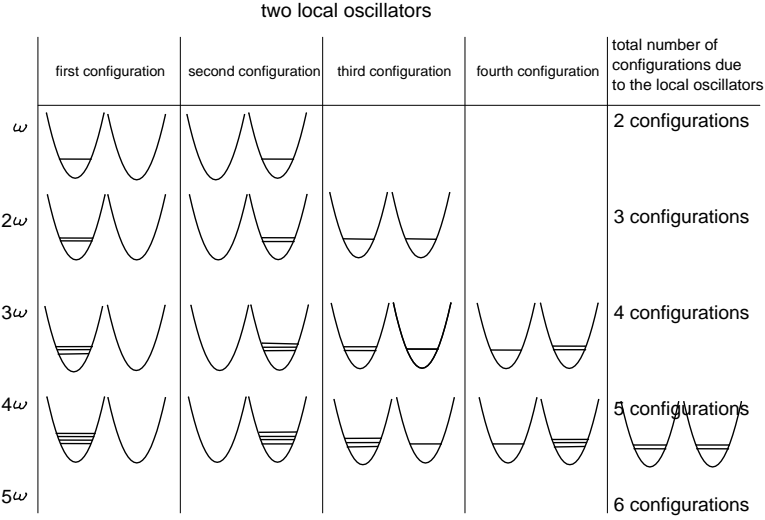}}
\end{minipage}
\end{center}
\caption{Local vibrational configurations on the energy shell of the initial wave packet
as a function of the excitation energy, hence as a function
of the temperature, available within the deformation resonance.
\label{phonon-counting}} \end{figure}

\noindent The dependence of the photon signal on the temperature is reflected by changes in the
vibrational structure of the local oscillators and the accessibility of the protein phonon configurations.
The temperature dependence of the accessible
vibrational configurations due to the local oscillators and the delocalized phonon configurations
in the protein is teated in the following way.
Figure \ref{phonon-counting} illustrates how the local oscillator vibrational structure changes with
temperature if we restrict the model to two local oscillators.
The wagging and the rocking modes are assumed to have the same energy
$\omega=5$ meV. This energy corresponds roughly to thermal
energy $k_BT$ at $T=58$ K. The energy is provided by the external heating
and is available within the deformation resonance.
So if the temperature in the 2D Fourier transform spectroscopy experiment equals 77 K, it means that
either one or the other local oscillator can be excited, i.e. there are two
degenerate vibrational configurations due to the local oscillators involved in the interaction
with the deformation resonance (the first line
in fig. \ref{phonon-counting}). At $T=125$ K and $T=150$ K
the available thermal energy is sufficient to excite just 2 local vibrations
which makes three degenerate local vibrational configurations
(the second line in fig. \ref{phonon-counting}) and so on.
Thus, when the temperature rises the local vibrational structure becomes more
complex. As the temperature rises 2, 3, 4 ... local vibrational configurations,
which are energetically accessible and are on the energy shell
with respect to the energy of the exciton wave packet, will be involved.
More and more on-shell local channels get open for the wave packet when the
temperature rises. This is the decisive feature of the model, which
prevents the destruction of the coherent time development by decoherence
and the total decay in the protein phonons. \\

\noindent Because we solve Schr\"odinger's equation with a procedure similar to CI,
we see that the number of the configurations involved increases with the
increasing energy, i.e. with increasing temperature.\\

\noindent The channels for decay in delocalized protein phonons also
increase, the avaibale density of phonon configurations
on the energy shell of the initial wave packet increases linearly with temperature.
This is so because we assume a one-dimensional protein chain with
constant density of states which provides a linear dependence on the temperature of the
on-shell density of phonon configurations.
If we would take the $3D$ Debye model for the phonons the density of on-shell phonon configurations
would vary with $T^3$, which would result in fast dissipation in phonons,
accompanied by the decay of the electron in the delocalized escape states. \\

\noindent The higher the temperature, the more protein phonon configurations are
accessible on the energy shell.
Assume the energy in the deformation resonance is 11 meV corresponding
to a temperature of 125 K. The phonon configurations in the protein backbone
which preserve the energy of the total wave packet
can vary their energy upto 11 meV. Assume now that the energy
in the deformation resonance is 22 meV corresponding to 250K.
The number of phonon configurations, which can preserve the
total energy of the wave packet and thus leave it on the energy shell,
is much larger.\\

\noindent The two channels, governed by the terms in the Hamiltonian:
\begin{eqnarray}
H_{el-CO}&=& ...
+V_{core-loc}c_v^+c_v\left( \sum_{i=1,2;n_i=0}^{\infty}d^+d_{n_i}+
\sum_{i=1,2;n_i=0}^{\infty}d_{n_i}^+d\right) \nonumber
\end{eqnarray}
and
\begin{eqnarray}
H_{el-phon}&=& Y_p \sum_{k}(c_v^+c_k +c_k^+c_v)\sum_{p}(b_p^++b_p) \nonumber
\end{eqnarray}
compete. The first term has the effect of conserving the amplitude of the wave packet in the local
region, whereas the second leads to its decay out of the local region in the
delocalized protein phonons, which are not measured in the experiment, causing
the exponential damping of the photon signal at short time.
As time goes by the two competing effects balance each other
and the photon signal gets stabilized at a finite value.\\

\noindent As the local structure in the model comprises more and more
configurations with rising temperature it has
a counterbalancing effect on the decay due to coupling to the delocalized phonons
and warrants that the weight of local configurations with the exciton
electron is retained locally. Then the third
laser pulse in the experiment can interact with the localized electron in the $3s$,
the deformation resonance and the local oscillators
and stimulate the photon emission with the amplitude retained in the local
structure.\\

\noindent
For our result the localized initial beable state is decisive.
In our theory we start from a local beable and follow its time development
during the tiny burst. Since the local structure expands
in number of configurations with rising temperature, the wave packet does
not delocalize completely and remains local. This reminds qualitatively of the
quantum Zeno effect, which is collapse after each interaction event with the environment.
However, the time development of the total system in high dimensional space
is completely coherent. The stochastic approaches try to obtain similar
effects by an empirical special choice and tuning of the phonon noise
such that the experimental beating signal is reproduced \cite{shabani}. \\

\section {Discussion}
\noindent As we see that with the simple model we reproduce
the experimental dependence of the photon amplitude on the waiting time and the
temperature, which is observed in the 2D Fourier transform electronic
spectroscopy, we can suggest an explanation of the questions
raised in the introduction.\\

\noindent No isotope effect on the coherences in FMO is established experimentally.
In our theory it is of no significance
if $^1H$ is replaced by $^2H$ or $^{12}C$ is exchanged with $^{13}C$. The local vibrational spectrum changes,
but we only need the spectrum on the energy shell of the wave packet.
The local picture explains why if something changes 10 {\rm \AA} away from
the transient negative ion resonance,
for instance isotopes of hydrogen are exchanged, the change does not affect the
coherent behaviour of the NIR, which is observed in experiment. \\

\noindent Using an argument from perturbation theory, the first order
coupling between two configurations scales with the inverse of their energy difference.
The closer the energy of the interacting configurations,
the stronger the interaction between them and the first order term
will provide the major contribution.
Hence, the on-shell contribution to the coupling dominates. Therefore the many particle
configurations we take into account, are nearly on-shell with the initial
wave packet since they provide the major contributions to the coupling.\\

\noindent Reminding the many-particle character of the field configurations
in our theory we necessarily have to conclude that the nature of the
states involved in the superposition, leading to the quantum beats,
is vibronic, as it has been suggested by experimentalists as well.\\

\noindent {\bf Why does conventional decoherence not prevail?}
The initial state in a conventional decoherence approach is the superposition of exciton
states delocalized over the protein cavity.
The environment are the protein phonons, the coupling
of the excitons with the environment leads to phonon excitation. Energy
conservation requires that the excitons are deexcited in the ground
electron state. No oscillations between the electron ground and exciton
states can occur. With the argument used by decoherence theory,
the reduced density matrix has then a single
non-zero diagonal matrix element equal to 1 for the electron
ground state of the excitons. \\

\begin{figure}{} \begin{center} \begin{minipage}{15cm}
\scalebox{0.425}{\includegraphics*{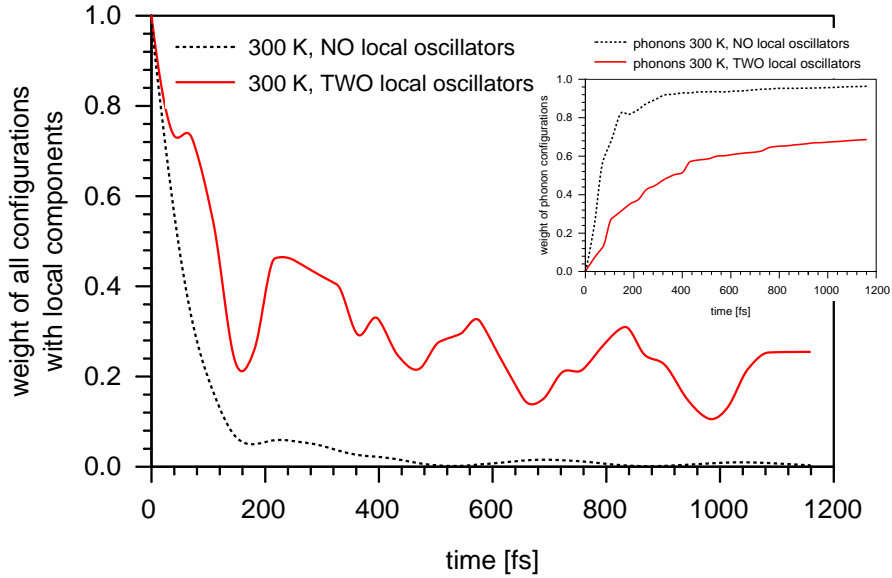}}
\end{minipage}
\end{center}
\caption{Time development of the photon amplitude in a model neglecting
the coupling of the exciton electron to the local oscillators with dominating coupling to the phonons
(decaying dashed curve), corresponding to temperature 300 K, compared to the result
with the local structure included (oscillating full curve). Inset: development with
time of the weight of field configurations involving the protein phonons
with the local structure included (full curve)
and excluded (dashed curve).
\label{engel-decoherence}} \end{figure}

\noindent This would happen if we would neglect the chooser
mechanism, which localizes the electron in the $3s$-affinity orbital
of the carbonyl carbon atom. In an attempt to simulate the result of decoherence theory let us
omit the local oscillators due to the carbonyl $CO$ fragment,
as in decoherence theory
they play no role. The initial state is also changed, the electron is no
more 'chosen' in the warp resonance, it is assumed to be
in the somewhat diffuser deformation resonance.
If in our model we would disregard the chooser mechanism and
the coupling to the local $CO$ oscillators and allow the exciton
electron to couple to phonon continua all over
the protein chain  the result is similar to what is expected from decoherence theory
(fig. \ref{engel-decoherence}).
The two curves in fig. \ref{engel-decoherence} are evaluated with different initial
states, of course. Whereas for the upper curve the exciton electron in the initial wave packet
is localized via the chooser mechanism in the beable, for the lower curve
where no chooser mechanism is operating, the initial wave packet has the electron
in the deformation resonance. Ignoring the chooser mechanism leads to
an exponential decay of the photon amplitude without oscillations to zero value,
completely and irreversibly.
Everything is lost in the protein phonons (cf. the dashed curve in the inset in fig. \ref{engel-decoherence}),
and this is not measured in the experiment.
This resembles the prediction of conventional decoherence theory.
In contrast, the chooser mechanism due to degeneracy entanglement with gravonons
determines the transient
localization of the electron wave packet
in a beable and precludes its total decay in the environmental phonon
continuum, which is the reason for retaining finite photon amplitude
for long time, at least within the lifetime of the tiny burst. \\

\noindent The curve, which reproduces the experimental result, starts with the exciton
electron in a beable (which is destroyed by the
photon-induced tiny burst) localized in the NIR via
the chooser mechanism due to entanglement with the gravonons.
The exponentially decaying photon amplitude curve is evaluated by neglecting the
chooser mechanism, the electron is in the deformation resonance, where it couples with
the dissipative phonon environment all over the protein backbone. \\

\noindent {\bf The role of the gravonons}, although they are not involved
during the time of the tiny burst, is to create the beable via the chooser mechanism.
The transient trapping of the exciton electron in the $3s$-affinity orbital on the carbonyl carbon atom
due to the entanglement
with the degenerate gravonons occurs before the double laser pulse
and the tiny burst. The laser pulse generates the tiny burst and causes
the bursting of the $3D$ components of the total wave packet in $3D$ space
and the destruction of the entanglement with the gravonons
(cf. fig. \ref{beable} and the inset showing how $3D$ components
of the wave packet gain weight during the tiny burst). \\

\noindent From our theory and the measurements it appears that {\bf decoherence is an illusion}.
If it did exist it would have led to the total decay of the photon signal within
femtoseconds, as the decaying dashed curve in fig. \ref{engel-decoherence} shows.
Some obvious arguments can be summarized:  \\
\noindent (i) Despite that in the present theory standard decoherence theory due to coupling
to environmental degrees of freedom is {\bf not} involved, all experimental observations are
reproduced in a coherent picture by solving the TDSE. Reduced
density operator and matrix, being the basis for the conclusions
in decoherence theory, are neither needed nor used.
The chooser mechanism localizes the exciton electron within the beable,
allowing it to interact with the local core movement states of the carbonyl group in
addition to the dissipative interaction with the phonons in the protein backbone.
In our theory these are physical phenomena within a completely coherent picture.\\
\noindent (ii) Decoherence theory misses the localization in a beable.
It starts from an initial state with a delocalized
exciton electron. In contrast,
in the present theory we reproduce and explain the experiments on the coherent behaviour of
the protein-FMO complex as effects of the local structure and its temperature dependence.
The generation of a local beable via entanglement
of the exciton electron with gravonons is the clue to the understanding of the experiment.
In contrast, decoherence theory has no local beables, therefore it cannot provide
the explanation of the long-living coherence of FMO.\\
\noindent (iii) In the stochastic approaches, based on the reduced density
matrix and hence decoherence, special constructions of environmental excitations, in
particular special phonons (colour-noise), are suggested as a decohering environment
to be able to reproduce the experimental curves.
Arguments are suggested concerning the efficiency of
electron transfer from the FMO complex to the reaction center.
But experiment shows the coherent behaviour
and {\bf no} dependence of the coherent behaviour on the protein phonon structure.\\
\noindent (iv) Stochastic approaches, missing localization in beables,
cannot provide an explanation of the non-zero
value of the photon echo amplitude at long time. No explanation is
suggested why should the phonons not be able
to reduce the photon echo amplitude to zero.
If decoherence were active, the large number of environmental degrees
of freedom involved should lead to complete decay of the
photon echo signal.

\section { Conclusion}
\noindent We can understand the experimental results
of long time coherence in the FMO protein complex without the far-fetched
assumption of special phonons involved in the approach based on stochastic quantum master equations.
Our understanding is based on the localization of the exciton electron
due to a localization chooser mechanism using gravitation in high dimensional
spacetime. Released from the entanglement with the gravonons
by the laser pulses, the
electron can couple with local vibrations of the carbonyl $CO$ group
and the delocalized phonons in the protein backbone.
This determines the vibronic character of the many-particle configurations
leading to the beating of the photon echo signal.
The two channels lead to preservation of a local non-vanishing
weight of the exciton electron and to the beating photon echo signal and its exponential damping.
Conventional decoherence theory would in fact
predict a total exponential decay of the signal
and much faster total loss of coherence compared to the
experimentally established coherence times.\\

\noindent In order to understand the finite final value of the photon echo signal in 2D
Fourier transform electronic spectroscopy
at long time we involve a very local structure and then
a temperature dependent accessibilty to the local structure.
The starting point in our theory on the coherent behaviour of chlorophyll assemblies is the localized exciton
electron as a beable which allows the interaction of this localized charge
with the photon field. The further time development of the wave packet
follows from the initial localization of the electron. \\

\noindent The stepwise bunching of photon-amplitude curves at different temperatures
(fig. 3 for $T=125$ K and $T=150$ K in ref. \cite{engeletal}) is only possible if we have a local picture with
excitation of only a few local oscillators. This point needs to
be further investigated experimentally since the error bars in the experimental figure
in ref. \cite{engeletal} are larger than the expected differences between the curves.\\

\noindent The observation that isotope substitution does not change the
experimentally measured photon echo amplitude is explained in our theory.
No special phonons ('colour noise', special spectral function)
are needed to explain the coherent behaviour and the
damped amplitude of the photon beating signal with time.\\

\noindent We explain the magic non-appearance of decoherence in the
2D Fourier transform electronic spectroscopic experiment. \\

\noindent To finalize, we claim that quantum mechanics needs a chooser mechanism for particle
localization. Experiments can be explained within quantum mechanics only
if the incoming delocalized particle wave gets localized on a local site.
The solution of the time dependent Schr\"odinger equation in high dimensional spacetime including
the entanglement to gravonons is the theory of the chooser mechanism.\\

\noindent Local beables are the fundament of measurement. The measurement
gives data on local features and this is what Schr\"odinger's equation with entanglement with the
gravonons in high dimensional spacetime ($11D$) provides. All measured particles
are local, they are local only when they entangle with gravonons as beables.\\

\noindent {\bf Aknowledgment:} We gratefully aknowledge the useful discussion
and comments by G. Engel at the Gordon Center for Integrative Science,
Chicago University.

\end{document}